%% file: main.tex
\documentclass[
reprint,
superscriptaddress,
amsmath,amssymb,
aps,
prl,
floatfix
]{revtex4-2}

\usepackage{graphicx}
\usepackage{dcolumn}
\usepackage{bm}
\usepackage{wasysym}
\usepackage[colorlinks,citecolor=magenta]{hyperref}
\usepackage[T1]{fontenc}

\begin{document}

\title{Thickness Insensitive Nanocavities for 2D Heterostructures using Photonic Molecules}

\author{Peirui Ji}
\affiliation{State Key Laboratory for Manufacturing Systems Engineering, Xi’an Jiaotong University, Xi’an, 710049, China}
\affiliation{Walter Schottky Institut and Physik Department, Technische Universit{\" a}t M{\" u}nchen, Am Coulombwall 4, 85748 Garching, Germany}
\author{Chenjiang Qian}
\email{chenjiang.qian@wsi.tum.de}
\affiliation{Walter Schottky Institut and Physik Department, Technische Universit{\" a}t M{\" u}nchen, Am Coulombwall 4, 85748 Garching, Germany}
\author{Jonathan J. Finley}
\email{finley@wsi.tum.de}
\affiliation{Walter Schottky Institut and Physik Department, Technische Universit{\" a}t M{\" u}nchen, Am Coulombwall 4, 85748 Garching, Germany}
\author{Shuming Yang}
\email{shuming.yang@mail.xjtu.edu.cn}
\affiliation{State Key Laboratory for Manufacturing Systems Engineering, Xi’an Jiaotong University, Xi’an, 710049, China}


\begin{abstract}
  Two-dimensional (2D) heterostructures integrated into nanophotonic cavities have emerged as a promising approach towards novel photonic and opto-electronic devices.
  However, the thickness of the 2D heterostructure has a strong influence on the resonance frequency of the nanocavity.
  For a single cavity, the resonance frequency shifts approximately linearly with the thickness.
  Here, we propose to use the inherent non-linearity of the mode coupling to render the cavity mode insensitive to the thickness of the 2D heterostructure.
  Based on the coupled mode theory, we reveal that this goal can be achieved using either a homoatomic molecule with a filtered coupling or heteroatomic molecules.
  We perform numerical simulations to further demonstrate the robustness of the eigenfrequency in the proposed photonic molecules.
  Our results render nanophotonic structures insensitive to the thickness of 2D materials, thus owing appealing potential in energy- or detuning-sensitive applications such as cavity quantum electrodynamics.
\end{abstract}

\maketitle

\section{\label{sec1}Introduction}

Photonic crystals have spatially periodic dielectric constants that facilitate the engineering of photonic bandstructure via geometry \cite{PhysRevLett.58.2059}.
By introducing defects into the periodic structure, photonic confinements can be generated for cavities, waveguides and directional couplers \cite{10.1038/386143a0,10.2307/j.ctvcm4gz9}.
A single-mode photonic crystal cavity with a mode volume of the order of $\sim(\lambda/n)^3$ can be used to enhance the strength of light-matter coupling towards the quantum limit \cite{RevModPhys.87.347}, whilst allowing individual components to be combined into integrated photonic circuits.

The conventional fabrication methods used to realize photonic crystals is to etch a periodic array of air holes in a high-index semiconductor slab such as Si and GaAs \cite{10.1038/383699a0,10.1002/lpor.201500321,10.1109/JPROC.2018.2853197}.
By improving the design \cite{10.1038/nature02063} and fabrication methods \cite{10.1016/j.photonics.2009.10.004}, such conventional photonic crystal cavities have been optimized towards the theoretical limit.
For example, Q-factors have been pushed into ultra-high ($>10^5$) regime, and the experimentally measured performance can reach the theoretical calculations \cite{10.1038/nmat1320}.
Such a cavity is ideal to study and control light-matter interactions by directly embedding quantum emitters such as quantum dots \cite{10.1038/nature03119,PhysRevB.71.241304,10.1038/nature05586} into the dielectric structure.
This type of cavity--quantum-dot system has been widely applied in quantum devices \cite{10.1038/nphys343,10.1038/nphoton.2013.48}, and extended to study multi-modal light-matter interactions such as multi-photon processes \cite{PhysRevLett.107.233602,PhysRevLett.120.213901}, electromagnetic controllability \cite{PhysRevLett.103.127401,PhysRevLett.104.047402,PhysRevLett.122.087401} and phononic effects \cite{PhysRevB.80.201311,10.1063/1.4824712}.

However, conventional photonic crystal cavities are less suitable for integration with 2D atomically thin semiconductors.
This is since it is non-trivial to embed 2D semiconductors into the dielectric slab during growth, their properties are strongly influenced by the dielectric disorder, and encapsulation with an ultra-flat, inert insulator is essential for high-performance 2D devices \cite{10.1038/s41598-017-09739-4,PhysRevX.7.021026,10.1038/s41565-019-0520-0,10.1038/s41378-021-00332-4,10.1002/adma.202203889}.
Hexagonal boron nitride (hBN) is a layered 2D insulator and has proven to not only be an ideal substrate for other 2D materials, but also host a variety of defect quantum emitters having potential interests for nanophotonics.
The integration of hBN in photonic crystals raises new challenges in the nanofabrication \cite{10.1038/s41467-018-05117-4,10.1002/adom.201801344}.
Meanwhile, in contrast to conventional cavity materials such as Si and GaAs whose thickness can be precisely controlled during growth, the thickness of 2D materials is difficult to control and even accurately identify.
The indeterminate thickness of 2D heterostructures will result in shifts in the frequency of 2D-material cavity from the designed value \cite{10.1038/s41467-018-05117-4,10.1002/adom.202200538}.
Recently, we have proposed a hybrid nanobeam cavity to solve the problem of etching of 2D heterostructures \cite{PhysRevLett.128.237403,PhysRevLett.130.126901}, but the problem of 2D-material thickness fluctuation has not been fully investigated.

\begin{figure*}
  \includegraphics[width=\linewidth]{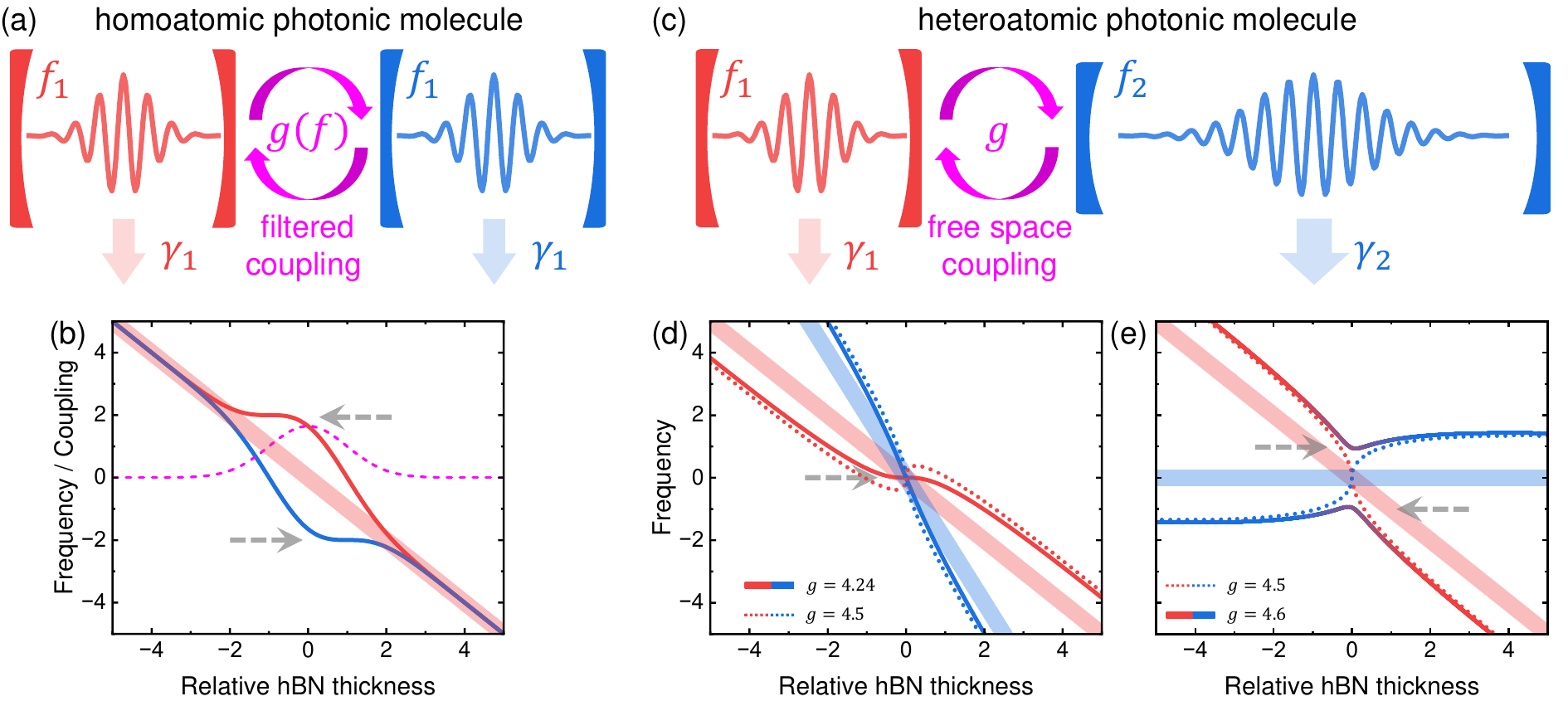}
  \caption{\label{f1}
    Coupled mode theory.
    (a) Schematic of the homoatomic photonic molecule.
    Two cavities have the same frequency $f_1$ and loss rate $\gamma_1$.
    A frequency-dependent (filter) coupling between them is used to achieve a parameter regime having a robust eigenfrequency.
    (b) The light line represents the bare frequency $f_1$.
    The purple line is the frequency-dependent coupling strength $g(f)$.
    The solid lines represent the eigenfrequencies of the molecule.
    (c) Schematic of the heteroatomic molecule.
    Two cavities have different frequencies $f_{1(2)}$ and loss rates $\gamma_{1(2)}$.
    (d)-(e) The light lines represent the bare frequencies $f_{1(2)}$.
    The dotted lines are the eigenfrequencies at the diabolical point (threshold between weak and strong coupling).
    Solid lines represent the eigenfrequencies of the molecules having a robust regime.
    Gray dashed arrows in (b) and (d)-(e) denote the position of robust eigenfrequencies (region with vanishing slope).
  }
\end{figure*}

Here, we report several approaches to realize a robust resonance frequency with 2D heterostructures using photonic molecules.
Such a photonic molecule consists of two coupled cavities and usually host coupled modes with spatially symmetric (S) and antisymmetric (AS) character arising from the mode coupling \cite{10.1063/1.3475490,10.1364/OE.21.016934,10.1364/OE.23.009211,10.1364/AOP.376739}.
For a single cavity, the resonance frequency generally shifts linearly with the thickness of the 2D heterostructure.
Therefore, we exploit photonic mode coupling to achieve nonlinear shifts of the eigenfrequencies of coupled cavities.
In this case, the linear slope of frequency shift is suppressed to zero at the local inflection points, enabling a resonance frequency robust to the thickness of the 2D materials.
Based on the coupled mode theory, we find that such nonlinear eigenfrequencies can be achieved either using a homoatomic molecule with frequency filtered coupling or using heteroatomic molecules around the diabolical point.
These predictions are further demonstrated quantitatively using 2D and 3D finite-difference time-domain (FDTD) calculations.

\section{\label{sec2}Coupled Mode Theory}

For a cubic cavity, the wave vector of the cavity mode is proportional to the inverse of the spatial extent, and thereby, the resonance frequency monotonically decreases with the increasing cavity size.
This is generally the same to a single nanocavity based on conventional photonic crystal structures.
We consider the bare frequency $f_1$ of a single cavity that varies linearly with the hBN thickness $t_{hBN}$.
Thus, the aim is to construct a nanophotonic molecule with eigenfrequencies $f_\pm$ that vary non-linearly and nonmonotonically with $t_{\mathrm{hBN}}$.
As such, the first order derivative (linear slope) of $f_\pm$ to $t_{\mathrm{hBN}}$ is zero at the minima or maxima points, which means the resonance frequency is robust within a range of hBN thickness.

We consider two individual cavities with the bare frequency $f_{1(2)}$ and the photon loss rate $\gamma_{1(2)}$.
The molecule is described using the coupled mode theory \cite{10.1109/5.104225} according to
\begin{eqnarray}
  \label{eqr}
  \frac{1}{2\pi}\frac{da_1}{dt} &=& -if_1a_1-\frac{\gamma_1}{2}a_1-iga_2 \nonumber \\
  \frac{1}{2\pi}\frac{da_2}{dt} &=& -if_2a_2-\frac{\gamma_2}{2}a_2-iga_1
\end{eqnarray}
where $a_{1(2)}$ is the field amplitude of two cavities and $g$ is the coupling strength between them.
Then the resonance frequencies $f_\pm$ are calculated by solving the eigenstates of Eq. \ref{eqr} as
\begin{eqnarray}
  \label{eqe}
  f_\pm=\frac{f_1+f_2}{2}\pm \mathrm{Re}\left(\sqrt{g^2+\left(\frac{\Delta f- i\Delta \gamma/2}{2}\right)^2}\right)
\end{eqnarray}
where $\Delta f=f_1-f_2$ and $\Delta \gamma=\gamma_1-\gamma_2$.
For brevity, in this work, we define all parameters in frequency units, i.e., the coupling strength is $hg$ when converted to the energy unit where $h$ is the Planck constant.
For the results of coupled mode theory based on Eq. \ref{eqr} and \ref{eqe}, we omit the unit THz for brevity.

By solving Eq. \ref{eqr} and \ref{eqe}, we obtain several cases having nonlinear and nonmonotonic $f_\pm$, as presented in Fig. \ref{f1}.
We set cavity C1 (denoted by red) the frequency to $f_1=-t_{\mathrm{hBN}}$ and the loss rate to $\gamma_1=2$, and take it as the one that we want to suppress the sensitivity of the frequency to the thickness.
As schematically shown in Fig. \ref{f1}(a), we first consider a homoatomic molecule, meaning that the cavity C2 (denoted by blue) is the same as C1.
In this case, the two eigenfrequencies are $f_\pm=f_1\pm g$, with a splitting of $2g$.
Therefore, to achieve a nonlinear $f_\pm$ we need a nonlinear $g$.
One possible way is to introduce an optical filter \cite{10.1063/1.1634373,10.1007/s11107-017-0732-x}, which allows only the photons within certain frequencies to exchange between the two cavities.
Then it is expected that the coupling factor $g$ is frequency dependent, e.g., following a Gaussian filter as
\begin{eqnarray}
  g\left(f\right)=g_0\cdot \mathrm{exp} \left({-\frac{(f - f_0)^2}{2\sigma_g^2}}\right)
\end{eqnarray}
where $\sigma_g$ is the bandwidth of the filter and $g_0$ is the coupling strength at $f_0$.
In this case, we obtain that when $g_0=\sqrt{e}\sigma_g$, eigenfrequencies $f_+$ at $t=f_0-\sigma_g$ and $f_-$ at $t=f_0+\sigma_g$ have a zero value for both first-order (linear slope) and second-order derivative which means robust eigenfrequency regimes.
Fig. \ref{f1}(b) presents the eigenfrequencies with $g_0=1.648$, $f_0=0$ and $\sigma_g=1$ for example.
The purple line is $g(f)$, the light red line is the bare frequencies of single cavity $f_1$, and the solid red and blue lines are the eigenfrequencies $f_\pm$.
As shown, there are regions with robust $f_+$ or $f_-$, denoted by the gray dashed arrows.

The homoatomic molecule depicted schematically in Fig. \ref{f1}(a) requires a filtered photonic mode coupling.
In contrast, direct coupling in the free space between two cavities is much easier to achieve from a technological perspective.
In this case, the coupling strength $g$ is determined by the separation of the two cavities, whilst the frequency dependence is not significant.
As discussed above, if $g$ is nearly constant, the eigenfrequencies $f_\pm$ of a homoatomic molecule will still vary linearly.
Therefore, we construct heteroatomic molecules to achieve nonlinear $f_\pm$ in the free space, as schematically shown in Fig. \ref{f1}(c).
The cavity C1 (red) is identical to that used in the homoatomic case which has $f_1=-t_{\mathrm{hBN}}$ and $\gamma_1=2$.
We set the cavity C2 (blue) as a lower Q-factor with the larger loss rate $\gamma_2=20$.
Eq. \ref{eqr} and \ref{eqe} indicate that, when the coupling strength exceeds the decay rate $g>\Delta \gamma/4$, the molecule operates in the strong coupling regime with a splitting.
In contrast, in the weak coupling regime $g<\Delta \gamma/4$, it operates without an anticrossing \cite{10.1364/OL.37.003435}.
Moreover, the eigenfrequencies $f_\pm$ exhibit strong non-linearity and non-monotonicity around the diabolical point $g=\Delta \gamma/4$, the degenerate state at the threshold between weak and strong coupling \cite{10.1088/0305-4470/36/8/310,10.1038/s41377-020-0244-9}.

Fig. \ref{f1}(d) shows the calculated eigenfrequencies with $f_2=-2 t_{\mathrm{hBN}}$, meaning that the low-Q cavity C2 shifts more rapidly than the high-Q cavity C1 with the hBN thickness.
The light lines (red and blue) are the bare frequencies $f_{1(2)}$, whilst the dotted lines (red and blue) are the eigenfrequencies $f_\pm$ at the diabolical point $g=4.5$.
As shown, the high-Q mode (red dotted line) exhibits strong non-linearity and non-monotonicity.
In contrast, the solid lines (red and blue) are the case of $g=4.24$ which is slightly below the diabolical point.
Since the system is in a weak coupling regime, the superposition of the two cavities is not significant, and thereby, cavity C1 dominates in the spatial profile of the high-Q mode.
Moreover, in this case, the first-order (linear slope) and second-order derivative of the high-Q mode frequency (red line) to $t_{\mathrm{hBN}}$ are both zero at the resonance $t_{\mathrm{hBN}}=0$, thus generating robust eigenfrequency regimes denoted by the gray dashed arrows in the figure.

Fig. \ref{f1}(e) presents the results with $f_2=0$, which means the low-Q cavity C2 is not an hBN cavity.
The light lines (red and blue) are the bare frequencies $f_{1(2)}$, whereas the dotted lines (red and blue) represent the situation at the diabolical point $g=4.5$.
For $f_2=0$, we cannot use a $g$ which is slightly below the diabolical point, because the high-Q mode (red) blue shifts at $t_{\mathrm{hBN}}<0$ whilst redshifts at $t_{\mathrm{hBN}}>0$.
As a result, the indeterminacy (slope) around the resonance is enhanced rather than suppressed.
In contrast, the robust eigenfrequencies are achieved by $g$ slightly above the diabolical point, as shown by the case of $g=4.6$ (solid red-blue lines) in Fig. \ref{f1}(e).
In this case, the eigenfrequencies exhibits strong non-monotonicities, which provides the minima and maxima points with zero linear slopes.
Meanwhile, the mixture of red and blue color around the resonance $t_{\mathrm{hBN}}=0$ denotes that the eigenstates are the superposition of modes from both two cavities, which means their spatial profile still involves cavity C1, the one we want to suppress the indeterminacy.
Therefore, the case of $g=4.6$ in Fig. \ref{f1}(e) is also valid to achieve robust resonance frequencies.

\section{\label{sec3}2D-Material Cavities and Mode Couplings}

\begin{figure*}
  \includegraphics[width=\linewidth]{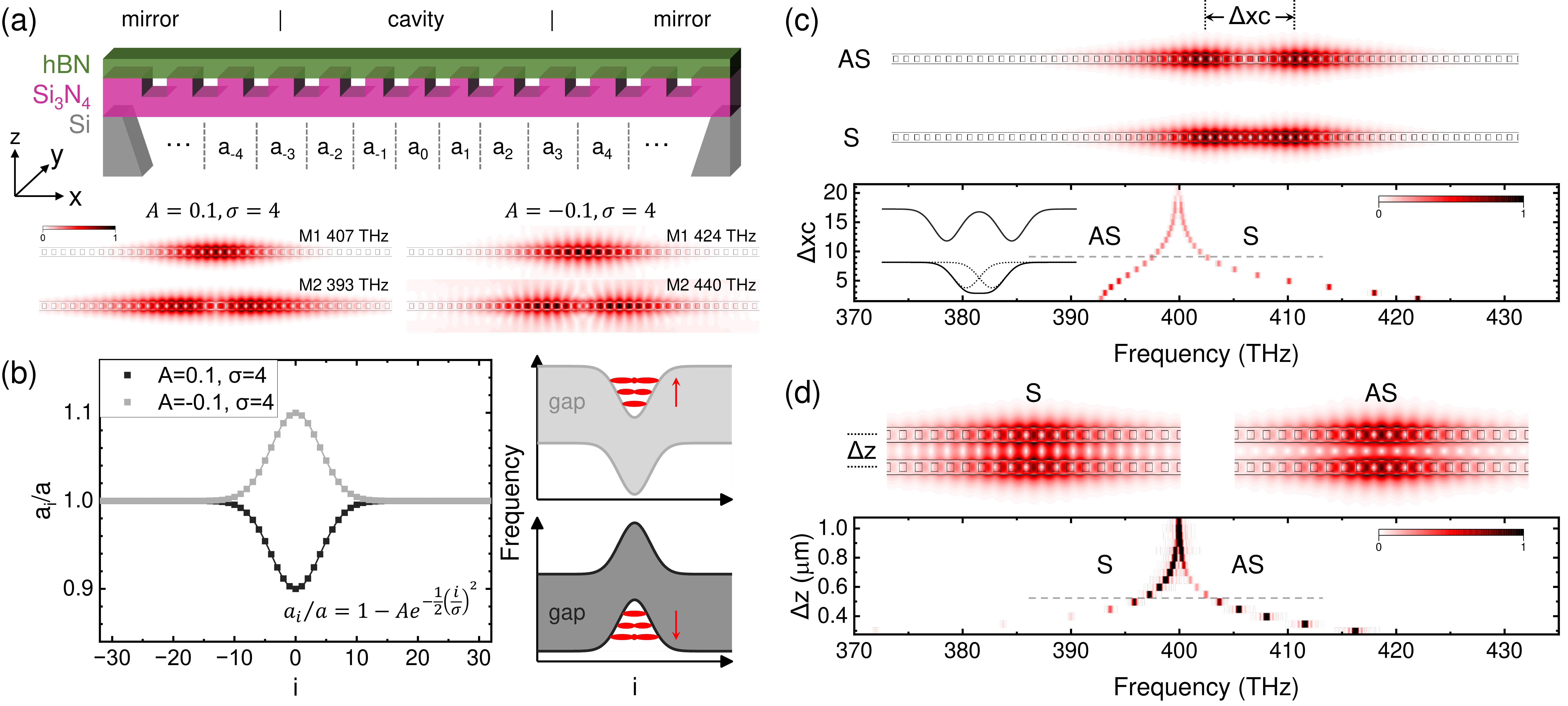}
  \caption{\label{f2}
    Fundamental properties of a single cavity and the molecule.
    (a) Schematic of the hBN/Si$_3$N$_4$ nanobeam cavity and typical photonic modes for a cavity with $A>0$ (left) or $A<0$ (right).
    (b) The photonic confinement is achieved by varying the periodicity at the center, defined by the Gaussian function with $A$ and $\sigma$.
    Insets show that determined by if $A$ is negative or positive, the confinement is concave or convex.
    (c) Coupling two cavities within one nanobeam or (d) two nanobeams.
    $\Delta xc$ in (c) and $\Delta z$ in (d) is the distance between the center point of two cavities.
    Gray lines and insets in the spectra in (c)(d) denote that the coupled mode theory is valid for a relatively large distance, i.e., $\Delta xc>8$ in (c) and $\Delta z >$ 500 nm in (d).
    In contrast, if the distance is too small, the two-photon confinements merge into one.
  }
\end{figure*}

The excitonic properties of 2D semiconductors are degraded when placed into an inhomogeneous environment such as dielectric disorder \cite{10.1038/s41565-019-0520-0}.
In this case, the exciton linewidth is broadened, and some novel single-photon emitters will be suppressed \cite{10.1038/s41467-019-10632-z,D0CS01002B}.
Therefore, 2D heterostructures with full hBN encapsulation is necessary to achieve the pristine excitonic and optical qualities \cite{10.1038/s41598-017-09739-4,PhysRevX.7.021026}.
Nowadays, the most common method used to build 2D heterostructures and integrate them into a nanocavity is dry viscoelastic stamping followed by transfer on top of pre-fabricated photonic cavities \cite{10.1038/nature14290,10.1038/nphoton.2015.197,10.1038/nnano.2017.128,10.1364/OME.443536}.
However, the attachment limits the cavity-matter overlap and decreases the cavity Q-factor.
Another way is to directly fabricate nanophotonic structures using hBN \cite{10.1038/s41467-018-05117-4,10.1002/adom.201801344}, with the problem that hBN cannot readily be etched \cite{10.1116/1.4826363}.
To solve these problems, we recently proposed a hybrid nanobeam cavity design \cite{PhysRevLett.128.237403,PhysRevLett.130.126901} that integrates hBN into nanocavities whilst avoiding etching of nanostructures into hBN layers.
As presented in Fig. \ref{f2}(a), all nanoscale trenches are etched in Si$_3$N$_4$ using mature fabrication technologies, whilst the hBN layer is not perforated.
Such nanocavities have been experimentally demonstrated to simultaneously achieve pristine excitonic quality, high Q-factor and large cavity-matter overlap, providing an ideal platform to explore the fundamental physics of 2D materials and their light-matter interaction in nanophotonic systems \cite{10.1021/acs.nanolett.2c00739,2210.00150,2302.07046}.

As presented in Fig. \ref{f2}(b), the cavity photonic confinement is formed by locally chirping the photonic crystal periodicity around the center point.
The distance between nanoscale trenches (periodicity) follows a Gaussian profile
\begin{eqnarray}
  \label{phc}
  a_i/a=1-A\mathrm{exp} \left({-\frac{i^2}{2\sigma^2}}\right)
\end{eqnarray}
where $a_i$ is the $i$th distance schematically shown in Fig. \ref{f2}(a), $a$ is the lattice constant, and $A,\ \sigma$ defines the Gaussian profile.
We fix the nanoscale trenches with the length $h_x=120$ nm, depth $h_z=150$ nm and lattice constant $a=$ 270 nm, whilst the Si$_3$N$_4$ has a thickness of 200 nm.
The designed hBN thickness $t_{\mathrm{hBN}}$ is 50 nm.
At the bottom of Fig. \ref{f2}(a), we present typical modes for two cavities with $A=0.1,\ \sigma=4$ and $A=-0.1,\ \sigma=4$ calculated using 2D FDTD.
For $A=0.1,\ \sigma=4$, the fundamental mode M1 is at 407 THz with a Q-factor of $4.8\times10^6$ and the second mode M2 is at 393 THz with a Q-factor of $1.8\times10^4$.
For $A=-0.1,\ \sigma=4$, the fundamental mode M1 is at 424 THz with a Q-factor of $5.9\times10^3$ and the second mode M2 is at 440 THz with a Q-factor of 820.
Since the refractive index of hBN is not well established \cite{10.1002/pssb.201800417,10.1364/OL.44.003797}, we use a fixed value of 2.0 for both hBN and Si$_3$N$_4$, while the mesh size used was 10 nm, providing a precision better than 0.01$\%$.

We observe that for $A>0$, the frequency of mode M2 is lower than M1, while it is higher when $A<0$.
This different mode frequency sequence arises from the different photonic bandgap as presented schematically in Fig. \ref{f2}(b).
For $A<0$, the periodicity at the center is increased.
The larger periodicity means a longer wavelength (lower frequency), thus the bandgap confinement is concave at the center.
Thereby, the higher order mode (larger mode volume) has a higher frequency.
Vice versa, for $A>0$ the bandgap confinement is convex, and thereby, the higher order mode has a lower frequency.
The cavity with $A<0$ has a lower Q-factor, and the field mainly distributes in the air gaps.
Therefore, the convex bandgap confinement is usually used in 1D nanobeam photonic crystal cavities \cite{10.1038/nnano.2017.128,10.1364/OME.443536}.

The photonic molecule can be constructed by coupling two cavities within one nanobeam or in two separated nanobeams.
For the case within one nanobeam, we define the periodicity profile as
\begin{eqnarray}
  \label{phcm}
  \frac{a_i}{a}=1 - A_1\mathrm{exp} \left({-\frac{(i-xc_1)^2}{2\sigma_1^2}}\right) - A_2\mathrm{exp} \left({-\frac{(i-xc_2)^2}{2\sigma_2^2}}\right) \nonumber
\end{eqnarray}
where the center point of the two cavities are at $xc_{1(2)}$ with the Gaussian profile $A_{1(2)}$ and $\sigma_{1(2)}$.
The coupling strength is controlled by the distance between them $\Delta xc=xc_1-xc_2$, as shown in Fig. \ref{f2}(c).
For the case with two separated nanobeams, we set $xc_1=xc_2=0$, and the coupling strength is controlled by the distance between two nanobeams $\Delta z$ as shown in Fig. \ref{f2}(d).
In Fig. \ref{f2}(c)(d) we present the distance-dependent spectra calculated using 2D FDTD, in the case of molecules with one or two nanobeams, respectively.
A homoatomic profile $A_{1(2)}=0.2$ and $\sigma_{1(2)}=2$ are used.
The symmetric (S) and antisymmetric(AS) modes have opposite frequency sequences in the two cases since the bandgap confinement in the $x$ direction along the nanobeam is convex for $A>0$ whilst concave in the $z$ direction.

The S and AS modes of the photonic molecule can be generally described by the coupled mode theory in Eq. \ref{eqr} and \ref{eqe}, but it is not strict.
For example, it predicts the homoatomic molecule with $f_1=f_2$ to have a symmetric splitting with eigenfrequencies $f_\pm=f_1 \pm g$ and the loss rate of S and AS modes remains $\gamma_1$.
However, in the FDTD calculation results, e.g., the case in the upper panel of Fig. \ref{f2}(c) which corresponds to $\Delta xc=9$, the Q-factor is $1.2 \times 10^5$ for AS mode while $5.3 \times 10^5$ for S mode.
This is due to the limitations of the coupled mode theory.
Firstly, in Eq. \ref{eqr} and \ref{eqe} we use a real $g$ to describe the coupling, but in reality, this term could also involve an imaginary part which results in a non-Hermitian system \cite{10.1364/AOP.376739,PhysRevA.105.013523}.
The imaginary coupling results in different loss rates of S and AS modes.
Secondly, the coupled mode theory is valid for individual cavities that couple to each other through the evanescent field.
When the distance between the two cavities is large, i.e., $\Delta xc>8$ in Fig. \ref{f2}(c) and $\Delta z >$ 500 nm in (d), we observe a nearly symmetric splitting which indicates that the two cavities are approximately individual.
In contrast, the splitting becomes asymmetric as the distance decreases, because the two confinements merge into one and the two cavities are no longer individual.
The comparison between individual and non-individual cases is schematically presented in the inset of Fig. \ref{f2}(c).
Nonetheless, the calculation results in Fig. \ref{f2}(c)(d) reveal that for both types of the photonic molecule, when the coupling strength $g$ is smaller than 2.5 (corresponding to $\Delta xc=9$ or $\Delta z =$ 550 nm), the eigenfrequencies split symmetrically thus can be well described by the coupled mode theory.

\begin{figure}
  \includegraphics[width=\linewidth]{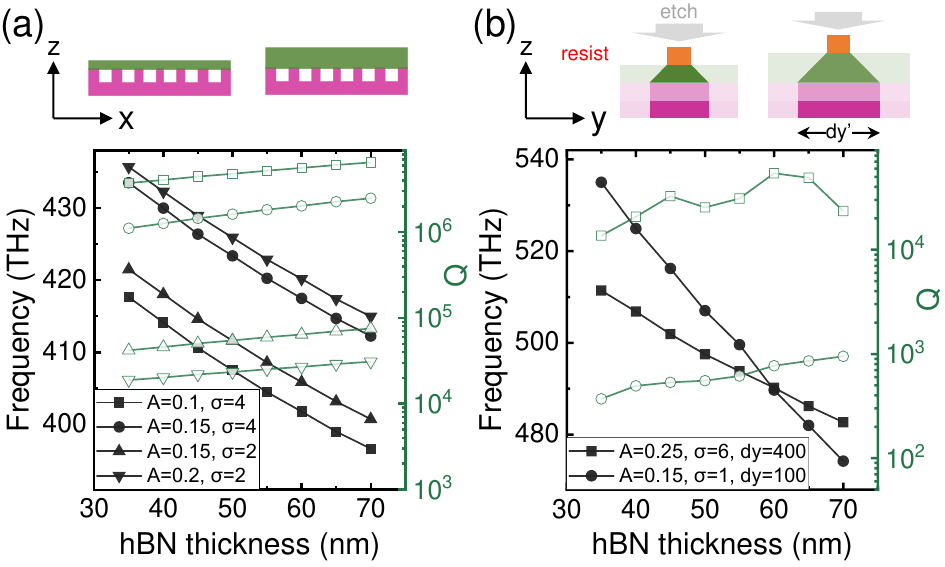}
  \caption{\label{f3}
    $t_{\mathrm{hBN}}$ dependence of a single cavity.
    (a) 2D FDTD calculation.
    For several values of $A$ and $\sigma$, the cavity mode exhibits a similar linear frequency shift.
    (b) 3D FDTD calculation.
    Due to the etching angle of hBN, $t_{\mathrm{hBN}}$ also affects the actual nanobeam width $dy'$.
    Thereby, we can control the frequency shift slope by the designed nanobeam width $dy$.
  }
\end{figure}

Before exploring the nonlinear eigenfrequencies for photonic molecules, we first study the hBN thickness dependence of the bare frequency of a single cavity.
Usually, apparent optical contrast is used to quickly identify the thickness of hBN flakes \cite{10.1063/1.4803041}, but the error of this method remains $\pm$ several layers, corresponding to a thickness of $\sim$ 2 nm.
In Fig. \ref{f3}(a) we present the hBN thickness ($t_\mathrm{hBN}$) dependence of cavity modes calculated using 2D FDTD.
For several different values of $A$ and $\sigma$, the cavity mode exhibits a similar frequency shift slope of -0.6 THz/nm.
This means that in the 2D space, it is difficult to have two cavities with remarkably different $t_\mathrm{hBN}$ dependent frequency shifts.
Therefore, to achieve the bare frequencies as predicted in Fig. \ref{f1}(d) of the two cavities, we extend the system to 3D space.
The dry etching of hBN normally has an etching angle that can be optimized by the recipe \cite{10.1116/1.4826363}, but it is difficult to make it completely vertical.
Due to the non-vertical characteristic, the hBN thickness $t_\mathrm{hBN}$ also impacts the nanobeam width $d_y'$.
In this work, we set the etching angle to be 45$^\circ$.
As schematically shown in the upper panel of Fig. \ref{f3}(b), for the same designed width $d_y$ (resist beam width), if $t_\mathrm{hBN}$ increases 5 nm, the actual nanobeam width $d_y'$ will increase by 10 nm.
The effect of increasing $d_y'$ from 100 to 110 nm is obviously different from that of 400 to 410 nm.
Therefore, we use a high-Q thick cavity with the designed nanobeam width $d_y$ of 400 nm, and a low-Q thin cavity with $d_y$ of 100 nm, to achieve remarkably different frequency shift slopes.
The frequencies and Q-factors of the two cavities calculated using 3D FDTD are presented in Fig. \ref{f3}(b).
As shown, the high-Q cavity with $A=0.25$, $\sigma=6$, $d_y=$ 400 nm has the frequency shift slope of $-0.8$ THz/nm, whilst the low-Q cavity with $A=0.15$, $\sigma=1$, $d_y=$ 100 nm has the slope of $-1.7$ THz/nm.

\begin{figure*}
  \includegraphics[width=\linewidth]{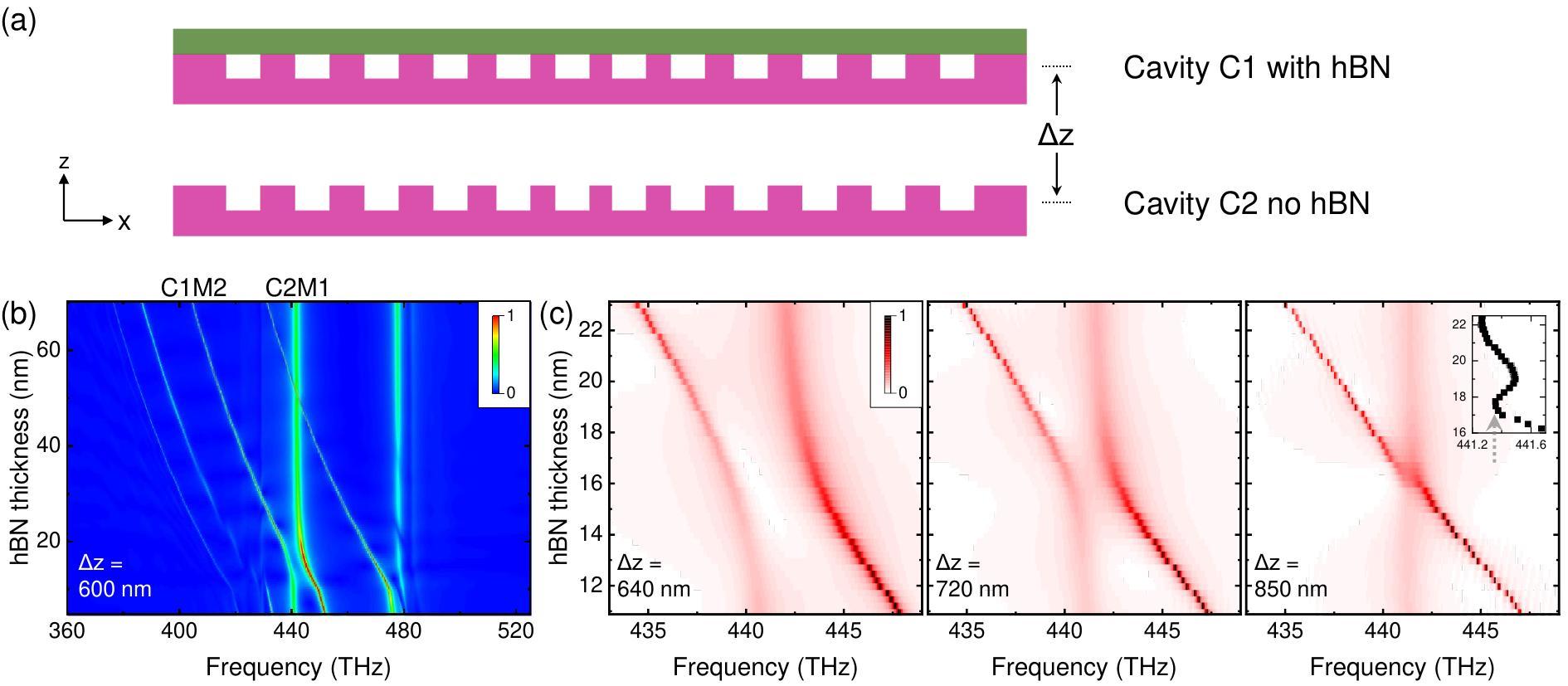}
  \caption{\label{f4}
    Robust eigenfrequency achieved by heteroatomic molecule involving one hBN cavity.
    (a) Schematic of the photonic molecule for which one cavity is with hBN whilst the other is not.
    The coupling between two cavities is controlled by the distance $\Delta z$.
    (b) Calculated spectra for the distance $\Delta z=$ 600 nm.
    (c) By increasing $\Delta z$, robust eigenfrequencies are observed at $\Delta z=$ 850 nm.
    Inset shows the frequency of AS mode which agrees well with the prediction in Fig. \ref{f1}(e).
  }
\end{figure*}

\begin{figure*}
  \includegraphics[width=\linewidth]{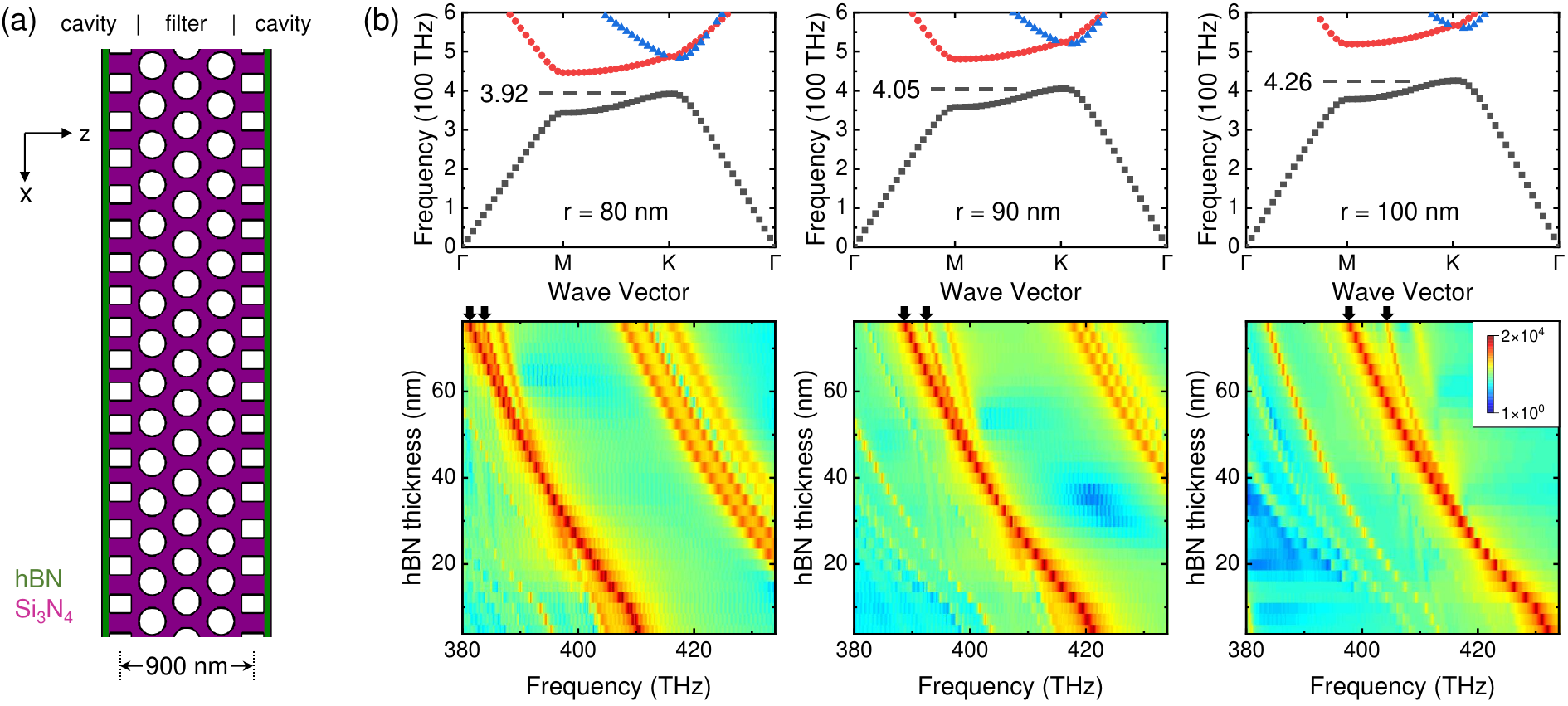}
  \caption{\label{f6}
    (a) Schematic of the photonic molecule with the filter coupler.
    (b) Upper panels are the bandstructure of the 2D hexagonal photonic crystal in the cases of different air hole radius $r$.
    Dashed lines denote the lower edge of the bandgap.
    The bottom panels are the corresponding spectra.
    Black arrows denote the split peaks.
    The splitting occurs above the threshold when its frequency reaches the photonic band shown in the upper panels.
  }
\end{figure*}

The 2D FDTD results in Fig. \ref{f2}(c)(d) reveal that the coupling strength of the photonic molecule can be controlled by the distance in $x$ and $z$ direction.
Similarly, the coupling can also be controlled by the distance in $y$ direction when we extend the system to 3D space \cite{PhysRevLett.128.237403}.
The 2D FDTD is a valid approximation of the 3D system, when the two cavities have the same nanobeam width $d_y$ in the $y$ direction.
Based on the results in Fig. \ref{f3}, we obtain that different $d_y$ in 3D space is only necessary to achieve remarkably different shifts of bare frequencies, which corresponds to Fig. \ref{f1}(d).
In contrast, uniform $d_y$ is enough to achieve the bare frequencies in Fig. \ref{f1}(b)(e).
Therefore, we use 2D FDTD to demonstrate the robust eigenfrequencies for the cases of Fig. \ref{f1}(b)(e) and 3D FDTD for the case of Fig. \ref{f1}(d).
From simple to complex, we next first show the case of Fig. \ref{f1}(e) in Fig. \ref{f4}, then the case of Fig. \ref{f1}(d) in Fig. \ref{f5} and finally discuss the case of Fig. \ref{f1}(b) in Fig. \ref{f6}.

\section{\label{sec4}Heteroatomic Molecules}

We firstly discuss the heteroatomic molecule comprising an hBN/Si$_3$N$_4$ hybrid cavity and a second cavity without hBN, corresponding to the case predicted in Fig. \ref{f1}(e).
Here we approximate the photonic molecule in the $xz$ 2D space as shown in Fig. \ref{f4}(a), and control the coupling $g$ between the two cavities by varying the distance $\Delta z$.
This approximation is of course not strict, but the key point, i.e., varying the distance between the two cavities to achieve a photonic molecule around the diabolical point, is the same for the real 3D structure and the 2D approximation.

We use $A=0.2,\ \sigma=4$ for the high-Q cavity C1 and $A=-0.3,\ \sigma=1$ for the low-Q cavity C2.
Since the coupled mode theory only applies to relatively large distance $\Delta z>$ 500 nm (Fig. \ref{f2}(d)), we first calculate the hBN thickness $t_\mathrm{hBN}$ dependent spectra at $\Delta z=$ 600 nm and present the result in Fig. \ref{f4}(b).
The four sharp peaks shifting significantly with $t_\mathrm{hBN}$ are from the high-Q cavity C1, and the two broad peaks exhibiting slight shifts are from the low-Q cavity C2.
As shown, the mode M2 of cavity C1 (C1M2) exhibits a clear anticrossing to the mode M1 of cavity C2 (C2M1).
The mode C1M2 has a bare frequency of 415 THz and a Q-factor of $2.5\times 10^4$ at the designed thickness $t_\mathrm{hBN}=$ 50 nm, and the mode C2M1 has a bare frequency of 441 THz and a Q-factor of 350.
Increasing the distance $\Delta z$ results in a reduction of the coupling strength $g$ between these two modes.
As presented in Fig. \ref{f4}(c), the splitting at resonance decreases with increasing $\Delta z$, and the system finally reaches the state slightly above the diabolical point as predicted in Fig. \ref{f1}(e).
The inset in Fig. \ref{f4}(c) shows that the resonance frequency of AS mode varies by 0.14 THz within the indeterminacy 2 nm of $t_\mathrm{hBN}$ ($17-19$ nm), one order of magnitude smaller than the case in a single cavity (1.2 THz).

\begin{figure}
  \includegraphics[width=\linewidth]{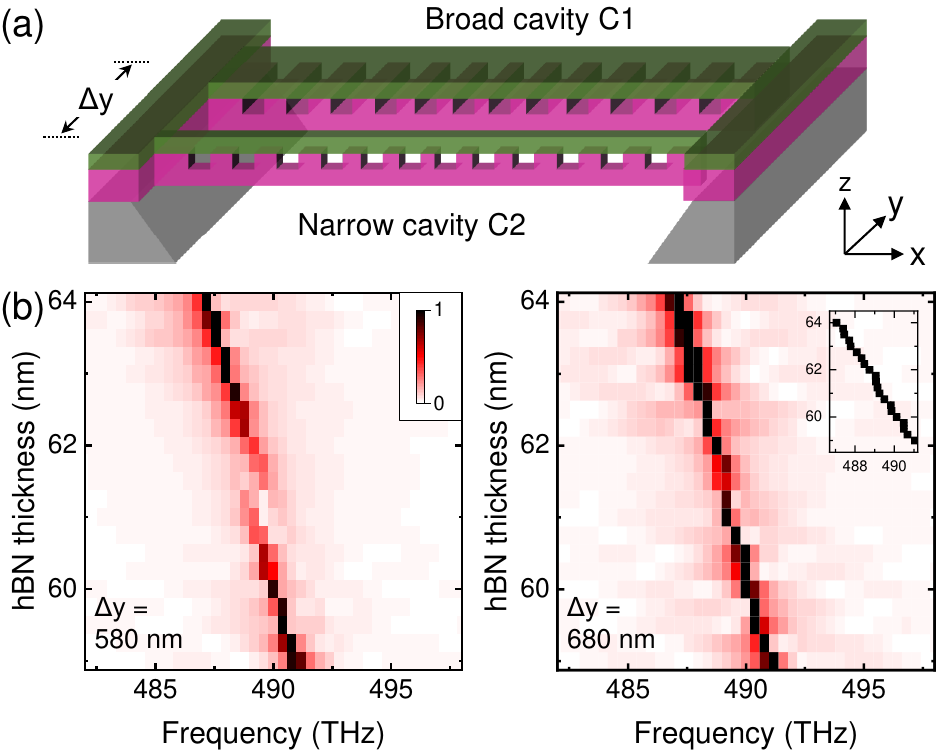}
  \caption{\label{f5}
    Robust eigenfrequency achieved by heteroatomic molecule consisting of two hBN cavities.
    (a) Schematic of the photonic molecule by coupling the two cavities in Fig. \ref{f3}(b).
    (b) Calculated spectra with the distance $\Delta y=$ 580 and 680 nm.
    Inset shows the frequency of high-Q mode at $\Delta y=$ 680 nm,  which agrees well to the prediction in Fig. \ref{f1}(d).
  }
\end{figure}

As discussed in the context of Fig. \ref{f3}(b), the frequency shift slope of a single cavity can be controlled by varying the nanobeam width $d_y$ in 3D space.
Therefore, we construct a photonic molecule based on these two cavities.
Hereby, we use $A=0.25$, $\sigma=4$, $d_y=$ 400 nm for the high-Q cavity C1 and $A=0.15$, $\sigma=1$, $d_y=$ 100 nm for the low-Q cavity C2.
The mode coupling is controlled by the distance $\Delta y$ as schematically shown in Fig. \ref{f5}(a).
In Fig. \ref{f5}(b) we present the spectra with different $\Delta y$ calculated using 3D FDTD.
The resonance point of the two cavities is at $t_\mathrm{hBN}=$ 61.5 nm, consistent with the bare frequencies of single cavities in Fig. \ref{f3}(b).
When $\Delta y=$ 580 nm, we observe a significant anticrossing indicating that the system is in the strong coupling regime.
In contrast, when $\Delta y=$ 680 nm, the coupling strength reduces and the anticrossing is suppressed.
Around the resonance $t_\mathrm{hBN}=$ 61.5 nm, the $t_\mathrm{hBN}$ dependent frequency shift of the high-Q mode is suppressed, as presented in the inset in Fig. \ref{f5}(b).
The 3D calculation results in Fig. \ref{f5} agree well with the state slightly below the diabolical point predicted in Fig. \ref{f1}(d).

\section{\label{sec5}Homoatomic Molecules}

In Fig. \ref{f1}(b) we predict that the homoatomic molecule with a filtered mode coupling $g(f)$ is another method to achieve nonlinear resonance frequency and improve the robustness to $t_\mathrm{hBN}$.
The optical filter is usually realized by integrating material with nonlinear absorption allowing only the photons at certain frequencies to pass.
However, the simulation of such complex materials in the FDTD calculation is non-trivial and we qualitatively explore the homoatomic molecule using a frequency selective coupler.
As discussed in the context of Fig. \ref{f3}, 2D FDTD is a valid approximation to describe the case of Fig. \ref{f1}(b), and thereby, we propose the nanophotonic device defined in the $xz$ plane as presented in Fig. \ref{f6}(a).
The two nanobeam cavities are located on each side, and an array of the 2D hexagonal photonic crystal lattice is in the middle.
Only the photons within the photonic band of the hexagonal photonic crystal are transmitted and exchanged between the two cavities.
Therefore, the hexagonal photonic crystal serves as a frequency-selective filter for the mode coupling.

We set $A=0.2,\ \sigma=4$ for the two nanobeam cavities.
The lattice constant of the hexagonal photonic crystal is set as $a_h=$ 270 nm, and we vary the radius of the air holes $r$ to control the photonic band.
The calculated photonic bands are presented in the upper panels in Fig. \ref{f6}(b), and the dashed lines denote the lower edge of the bandgap.
This value is 392, 405 and 426 THz for the radius $r$ of 80, 90 and 100 nm, exhibiting a positive dependence as expected.
The calculated $t_\mathrm{hBN}$ dependent spectra are presented in the bottom panels in Fig. \ref{f6}(b).
Multiple peaks are observed because the nanophotonic structure in Fig. \ref{f6}(a) is very complex.
Here we mainly focus on the peak with the largest intensity and its splitting, which are denoted by the black arrows on top of the spectra.
The splitting occurs when the hBN thickness $t_\mathrm{hBN}$ exceeds a threshold.
Moreover, this threshold in the cases of $r=$ 80, 90 and 100 nm corresponds to the frequency of 392, 402 and 415 THz, respectively.
These values are consistent with the band of the hexagonal photonic crystal denoted in the upper panels in Fig. \ref{f6}(b), indicating that the hexagonal photonic crystal is a valid filter for this photonic molecule.

We note that the nanophotonic structure in Fig. \ref{f6}(a) is a simple example to qualitatively explore the filtered coupling.
We use the distance of 900 nm because larger distance will suppress the coupling into the weak regime as shown in Fig. \ref{f2}(d).
As such, the hexagonal photonic crystal has only three rows of air holes, which is not enough for a highly efficient filter.
These limitations result in the relatively weak non-linearity, i.e., the $t_\mathrm{hBN}$ dependent shift of the splitting peak is reduced but not enough to obtain a robust (zero linear slope) resonance frequency.
A potential optimization approach is introducing $\Delta xc$ between the two nanobeam cavities.
In this case, we can control (through $\Delta xc$) the wave vector direction of the photons that exchange between the two cavities, and match the wave vector to some points in the $\mathrm{\Gamma-M-K-\Gamma}$ photonic band.
Thereby, the hexagonal photonic crystal will not only filter the frequency but also the wave vector, and thereby, the efficiency will be improved.

\section{\label{sec6}Conclusion}

In summary, we have demonstrated that the nonlinearity of the mode coupling in the photonic molecules results in coupled mode eigenfrequencies that are robust to the thickness of 2D heterostructure.
Although atomically thin semiconductors have well-defined emission energies compared to the random energy of traditional quantum dots, the indeterminate thickness of their hBN encapsulation leads to indeterminate cavity resonance frequencies.
For a single cavity, the frequency shift of 0.6 THz/nm is obtained in the 2D FDTD calculations, and a value of 1.7 THz/nm is obtained in the 3D calculations.
By comparison, the linewidth of the free excitons in the atomically thin semiconductors such as monolayer transition metal dichalcogenides at low temperatures is $\sim$5 meV (1.2 THz) and that of the localized excitons is $0.1-0.2$ meV ($\sim$50 GHz).
As a result, it is difficult to achieve the light-matter resonance directly in a single cavity, and complex external methods have to be used to control the detuning \cite{PhysRevLett.128.237403,2302.07046}.
Our results reveal that photonic molecules are capable of exhibiting thickness-independent resonance frequencies, thus having great potential in the 2D-material nanophotonics, especially the detuning sensitive applications such as lasing and exciton-photon polaritons.

\section*{Research Funding}

S. Y. and P. J. acknowledge support from the National Science Fund for Distinguished Young Scholars (52225507), the National Key Research and Development Program of China (No. 2021YFF0700402), and the Fundamental Research Funds for the Central Universities.
J. F., C. Q. and P. J. gratefully acknowledge the German Science Foundation (DFG) for financial support via grants SPP-2244, as well as the clusters of excellence MCQST (EXS-2111) and e-conversion (EXS-2089).
C. Q. gratefully acknowledges the Alexander v. Humboldt foundation for financial support in the framework of their fellowship programme.
P. J. acknowledges support from the China Scholarship Council (202006280067).

\section*{Author Contributions}
C. Q., J. F. and S. Y. conceived the project.
P. J. and C. Q. performed the calculations.
All authors discussed the results and wrote the manuscript.

\section*{Conflict of Interest}

Authors state no conflict of interest.

\section*{Data Availability}

The datasets generated during and/or analyzed during the current study are available from the corresponding authors on reasonable request.

\input{refer.bbl}

\end{document}

%% file: refer.bbl
%